\documentclass[pdflatex,sn-mathphys-num]{sn-jnl}

\usepackage{graphicx}%
\usepackage{multirow}%
\usepackage{amsmath,amssymb,amsfonts}%
\usepackage{amsthm}%
\usepackage{mathrsfs}%
\usepackage[title]{appendix}%
\usepackage{xcolor}%
\usepackage{textcomp}%
\usepackage{manyfoot}%
\usepackage{booktabs}%
\usepackage{algorithm}%
\usepackage{algorithmicx}%
\usepackage{algpseudocode}%
\usepackage{listings}%

\theoremstyle{thmstyleone}%
\theoremstyle{thmstyletwo}%

\theoremstyle{thmstylethree}%

\begin{document}

\title{Influence of the Ortho-II superstructure in the YBa$_2$Cu$_3$O$_{7-\delta}$ Orthorhombic phase after annealing}


\author*[1]{\fnm{Roberto F.} \sur{Luccas}}\email{luccas@ifir-conicet.gov.ar}
\author[2]{\fnm{Lorenzo} \sur{Gallo}}
\author[1,2]{\fnm{Cesar E.} \sur{Sobrero}}
\author[1,2]{\fnm{Jorge A.} \sur{Malarr\'{\i}a}}
%

\affil[1]{\orgdiv{Instituto de F\'{\i}sica Rosario}, \orgname{CONICET-UNR}, \orgaddress{\street{Bv. 27 de Febrero 210bis}, \city{Rosario}, \postcode{S2000EZP}, \country{Argentina}}}
\affil[2]{\orgname{Universidad Nacional de Rosario}, \orgaddress{\street{Pellegrini 1120}, \city{Rosario}, \postcode{S2000BTP}, \country{Argentina}}}


\abstract{Based on experimental results, this work proposes the influence of the Oxygen order present in the Ortho-II superstructure of YBa$_2$Cu$_3$O$_{7-\delta}$ (YBCO), on the final ordering of Oxygens in its Orthorhombic phase for $\delta$ $\approx$ 0.
Isothermal oxygenation of YBCO powder material is performed, starting from fully non-oxygenated material ($\delta$ $=$ 1) and evolving until saturation in an oxygen atmosphere.
The oxygenation process is carried out within a temperature range from 300 $^o$C to 800 $^o$C (300 $^o$C $<$ T$_O$ $<$ 800 $^o$C).
During the oxygenation process, and using a thermogravimetric balance, the evolution of mass (m) and the differential thermal analysis (DTA) of the material are monitored with respect to an inert reference material subjected to the same conditions as the YBCO powder (alumina powder).

These results allow observation of the Tetragonal-Orthorhombic (T-O) transition occurring in the YBCO material.
From these results, oxygenated YBCO material is obtained by working at different temperatures and under two different conditions: through a direct T-O transition into the Ortho-I superstructure, and by passing through the Ortho-II superstructure along the transition.
The material obtained under these two conditions is studied by X-Ray diffraction, revealing differences in the resulting diffractograms.
Furthermore, we propose that, for low values of T$_O$ (T$_O$ $<$ 400 $^o$C), the T-O transition proceeds through the region of the phase diagram where the Ortho-II superstructure is present, leading to progressive ordering of the Oxygen atoms within the material.
This ordering leaves a fingerprint in the final configuration reached by the YBCO material, even beyond the region where the Ortho-II superstructure is stable.
Finally, we suggest that this mechanism is responsible for the differences observed between the diffractograms obtained under both conditions.}




\maketitle

\section{Introduction}\label{SecIntro}

The YBa$_2$Cu$_3$O$_{7-\delta}$ (YBCO) compound, a flagship among high critical temperature cuprate-type superconductors, exhibits a strong interrelation between its superconducting properties and its Oxygen content.
Since the latter can be externally controlled, YBCO may therefore be regarded as an excellent candidate for externally tuning its superconducting properties, even on-the-fly.\cite{CupratesBM,ChuSC,LeeMott,LuccasArXiv}

Unlike other cuprates, YBCO presents one-dimensional Cu-O chains whose characteristics depend on the amount of Oxygen present in the material.
A given Oxygen content $\delta$ (0 $<$ $\delta$ $<$ 1) can diffuse throughout the material depending solely on the conditions to which it is subjected.
Consequently, this enables systematic exploration of its structural and electronic evolution from a fully non-oxygenated, non-superconducting state ($\delta$ $=$ 1) with tetragonal symmetry, to a highly oxygenated state ($\delta$ $\approx$ 0) with orthorhombic symmetry and superconducting properties.\cite{CavaOxYBCO,GunkelSmall25}

The Tetragonal-Orthorhombic (T-O) structural transition occurring during the oxygenation process of YBCO is closely related to the ordering of Oxygens incorporated into the Cu-O chains.
For intermediate values of $\delta$, this ordering does not necessarily lead to a single generic Orthorhombic phase based on full occupation of Oxygen sites along the Cu-O chains in the b-axis of the unit cell (the so-called Ortho-I superstructure).
Instead, it gives rise to a rich variety of superstructures characterized by periodic sequences of filled and empty chains (Ortho-II, Ortho-III, Ortho-V, and Ortho-VIII), alternating along the a-axis direction of the crystalline structure.\cite{deFontaineTeoric,TallonHole,Zimmermann03}
These superstructures reflect competition between long-range electrostatic interactions, frustration effects, and slow Oxygen kinetics; moreover, they play a fundamental role in charge transfer toward CuO$_2$ planes, thereby conditioning the superconducting properties of the material.\cite{Mook98,Tranquada95}

Detailed studies of these superstructures have primarily relied on highly sophisticated experimental techniques, such as high-energy X-Ray diffraction at synchrotron facilities and neutron scattering, which provide direct access to Oxygen ordering within the bulk.\cite{Zimmermann03,SchlegerOrto3}
In particular, extensive works have demonstrated that the Ortho-II, Ortho-III, Ortho-V and Ortho-VIII superstructures correspond to finite-range ordered states, with an effective dimensionality in the T-O phase diagram that depends on Oxygen content and temperature.
It has been observed that, although these superstructures generally do not develop long-range order, none of them exhibits three-dimensional ordering.\cite{Zimmermann03}

However, access to such experimental techniques is limited by their high cost and restricted availability, which constrains this approach for studying Oxygen dynamics and structural order in YBCO.
This, in turn, motivates the need for systematic studies aimed at determining whether structural signatures associated with the different Orthorhombic superstructures can be detected and analyzed using simpler and more accessible experimental techniques, such as conventional X-Ray diffraction, magnetic and electric measurements, and even complementary structural- and electronic-sensitive methods.\cite{LuccasArXiv,Veal90,Claus90}

In this work, we present an experimental study of the structure of YBCO material under Oxygen-saturation conditions ($\delta$ $\approx$ 0), performed at constant temperature for different oxygenating temperatures (T$_O$) above ambient temperature.
Based on previous results for this material,\cite{LuccasArXiv} different regions of the T-O phase diagram are explored.
The results reveal discrepancies between material oxygenated at low temperatures (T$_O$ $<$ 400 $^o$C, i.e., passing through the region with the Ortho-II superstructure) and material oxygenated at higher temperatures.
Accordingly, we propose the existence of a structural fingerprint associated with passage through the Ortho-II superstructure, persisting in the Oxygen-saturated state of the material, well within the Ortho-I ordered regime.
This conjecture is highly relevant, as it opens the possibility of controlling the Ortho-I structural state in YBCO, enabling the design of superconducting materials with tunable anisotropic, electronic, and dynamical properties.
Consequently, this broadens the spectrum of applications in sensor devices, microwave systems, and reconfigurable platforms based on superconducting materials, through in-situ modification of the chemical composition by altering only the value of $\delta$.\cite{GunkelSmall25}

\section{Results}\label{SecResult}

\begin{figure}
\begin{center}
\includegraphics[width=0.9 \columnwidth]{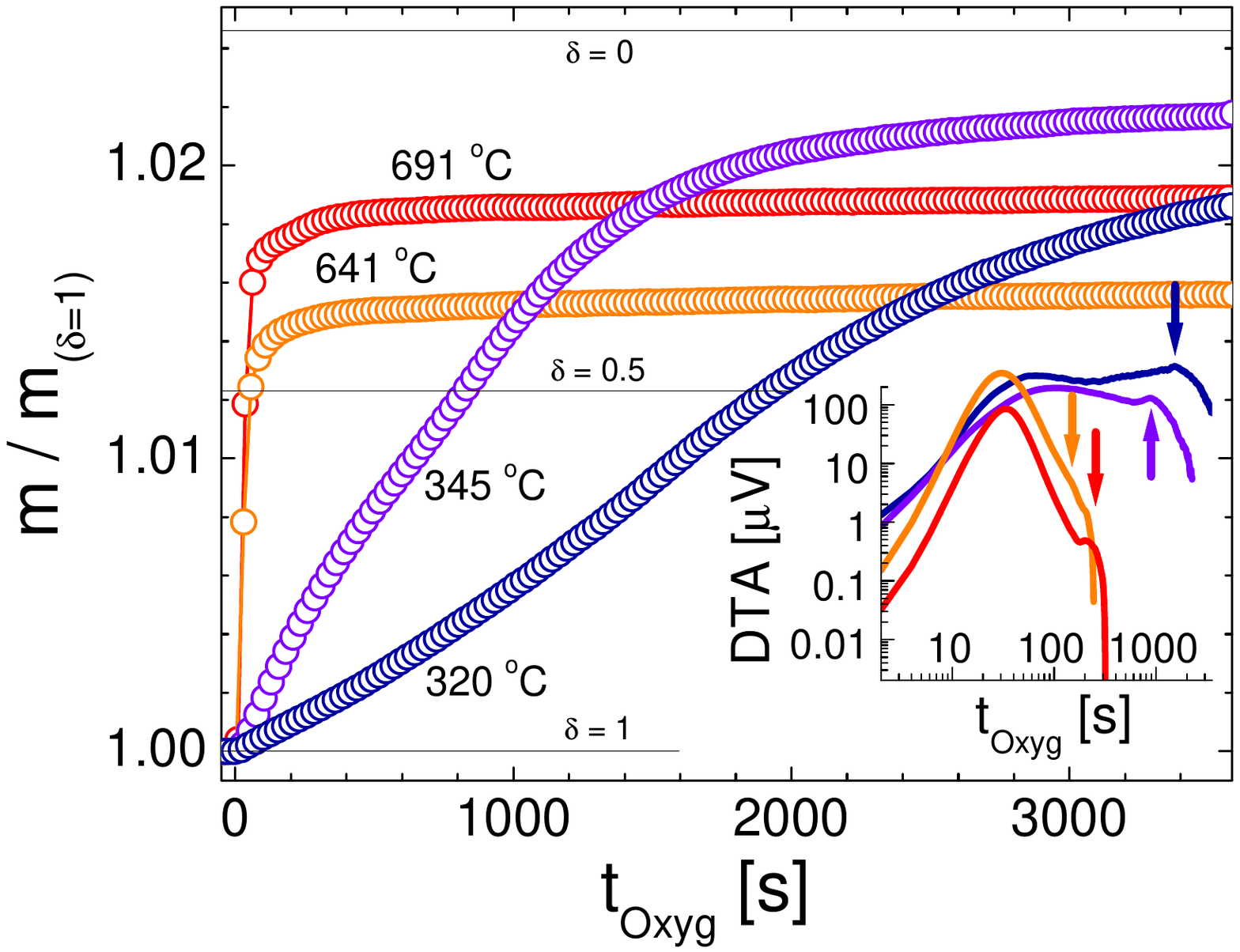}
\caption{\label{Fig1} 
Evolution of the mass {\emph{m}} for YBCO fully non-oxygenated ($\delta$ $=$ 1) during the annealing process at a constant temperature of oxygenation {\emph{T$_O$}}. 
Selected temperatures are shown as examples of typical evolution observed. 
Temperature values indicate the T$_O$ value for each set of data. 
As a guide to the eye, horizontal black lines mark several values of $\delta$ ($\delta$ $=$ 0, 0.5 and 1). 
The inset shows the related differential thermal analysis {\emph{DTA}} between the sample and an inert reference material (alumina). 
For better comprehension of DTA data, each set has been arbitrarily moved up with respect to the one for T$_O$ $=$ 691 $^o$C (data in red). 
Arrows indicate a second peak in each data set that is related to the T-O transition present in the material. 
This effect is also perceivable in the m/m$_{(\delta = 0)}$ as a change in the slope of the curves.}
\end{center}
\end{figure}

Experiments were carried out using YBCO powder of 5 $\mu$m of grain size as mentioned in Ref. \cite{LuccasArXiv}.
Thermogravimetric measurements were performed to follow the evolution of the mass (m) and the differential thermal analysis (DTA) of samples during oxygenation.
Fully non-oxygenated YBCO ($\delta$ $=$ 1) was used as starting material, and Oxygen saturation was reached in all the experiments performed.
Fig. \ref{Fig1} shows a typical evolution of m and DTA during oxygenation for different T$_O$ values.

The figure shows an increase in mass of the material upon exposure to it to an Oxygen saturated atmosphere, as well as saturation of Oxygen content at long oxygenation times t$_{Oxyg}$.
For the lowest temperatures shown, Oxygen saturation was reached only after more than five times the maximum t$_{Oxyg}$ displayed in the graph (t$_{Oxyg}$ $>$ 5 h).\cite{OBryan89}
In addition to mass increase, a change in the slope of m/m$_{\delta=1}$ prior to saturation is observed (much more clear for some T$_O$ values), indicating a change in the oxygen absorption dynamics.
This change in the oxygen absorption dynamics has been reported to be associated with the T-O transition in the material, due to reordering of Oxygen atoms within the Cu-O chains.\cite{LuccasArXiv}
In addition, the inset of the figure shows the DTA evolution, evidencing the exothermic nature of the oxygenation process.
In this inset, arrows indicate a second peak associated with the aforementioned slope change.
As noted, this second maximum arises from Oxygen reordering in the Cu-O chains due to the T-O transition.

\begin{figure}
\begin{center}
\includegraphics[width=0.9 \columnwidth]{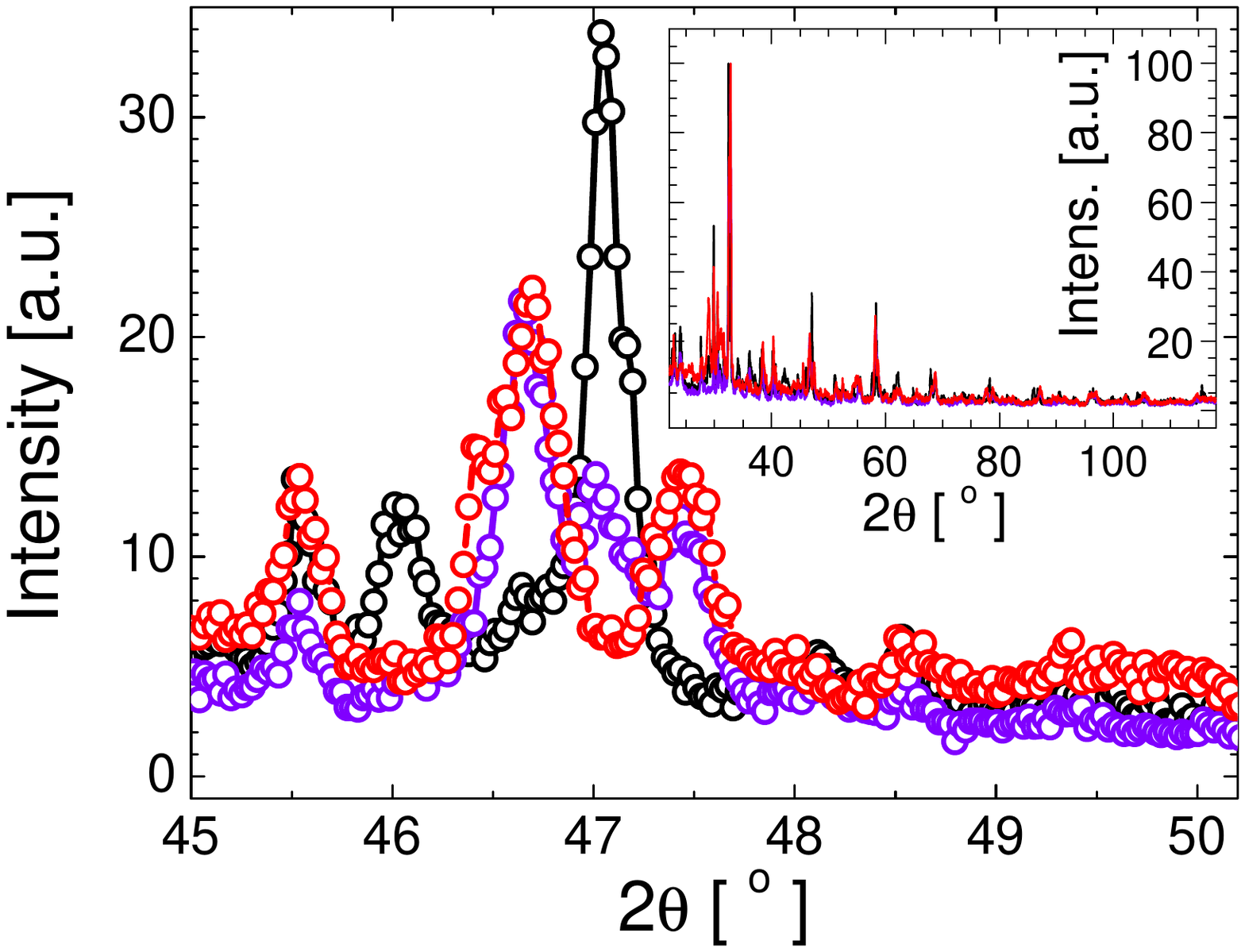}
\caption{\label{Fig2}
Typical diffractogram for YBCO in the Tetragonal (black) and in the Orthorhombic phases for T$_O$ below (violet) and above (red) 400 $^o$C. 
The inset shows the full scanned area, with a detail of one characteristic peak (2$\theta$ $\approx$ 47$^o$) in the main figure. 
Data shows the characteristic split of the peak at 47$^o$ when the T-O transition takes place.\cite{OBryan89} 
A difference between both difractograms performed in the Orthorhombic phase is present, since for low temperatures ($<$ 400 $^o$C) the peak at 47$^o$ does not disappear. 
Nevertheless, other peaks related to the Tetragonal phase (2$\theta$ $\approx$ 46$^o$) are not present for this particular orthorhombic case.}
\end{center}
\end{figure}

Consistent with previous results,\cite{LuccasArXiv} the oxygenation temperatures illustrated in Fig. \ref{Fig1} lie clearly within (T$_O$ $<$ 400 $^o$C) and outside (T$_O$ $>$ 400 $^o$C) the temperature range that runs across the zone of the T-O phase diagram where Ortho-II superstructure is present.
X-Ray diffraction experiments performed on these kinds of samples allow comparison of the material in both cases.
Fig. \ref{Fig2} shows typical diffraction results for the Tetragonal phase and for the Orthorhombic phase reached for T$_O$ below and above 400 $^o$C (i.e., after passing across the Ortho-II superstructure or not, respectively).
The figure presents a general view of the diffractograms for the three cases (inset), focused on a detailed view of one characteristic peak of the T-O transition.\cite{OBryan89}

The results show the typical peak splitting around the 2$\theta$ $=$ 47$^o$ position upon transition from the Tetragonal to the Orthorhombic phase, where one peak is present at 47$^o$ for the Tetragonal phase and two peaks are present at both sides of 47$^o$ for the Orthorhombic one.\cite{OBryan89}
However, results show here a clear difference between the two Orthorhombic phases under study (T$_O$ $>$ 400 $^o$C and T$_O$ $<$ 400 $^o$C).

In particular, for low oxygenation temperatures ($<$ 400 $^o$C), the splitting of the 47$^o$ peak occurs, giving rise to two new peaks on both sides.
However, the peak does not disappear; rather, it is still present.
On the other hand, at higher temperatures ($>$ 400 $^o$C), the 47$^o$ peak vanishes and it is replaced by the two peaks that characterize the Orthorhombic phase at both sides of the vanished one.
In addition, the peak at 46$^o$ associated with the Tetragonal phase disappears for all the experiments performed on the material in the Orthorhombic phase.

The absence of a full pattern of the Tetragonal peaks in the X-Ray diffractogram of the sample in the Orthorhombic phase obtained at low temperatures, suggests that there is no partial transformation of the material into the Orthorhombic phase.
This is due to the persistence of the 47$^o$ peak at low temperatures but not others like the one at 46$^o$, suggesting that no mixed diffraction pattern is present.
Moreover, in the Orthorhombic phase at low temperatures, where the 47$^o$ peak related to the Tetragonal phase persists, the system exhibits a high oxygenation level ($\delta$ $\approx$ 0.07), indicating a very low probability of remaining material in a Tetragonal phase.

\section{Discussion}\label{SecDisc}

From what has been said above, it is unclear whether the peak at 47$^o$ characteristic of the Tetragonal phase and observed only in the Orthorhombic phase obtained at low T$_O$, arises from residual material still in the Tetragonal phase.
The possibility that the corresponding $\delta$ fraction of the material remains entirely in the Tetragonal phase would imply a probabilistically unlikely ordering, given theoretical predictions that Oxygen vacancies within Cu-O chains repel each other.\cite{Aligia92PC190}
Moreover, considering the nature of samples (powder), that option is even harder to achieve.\cite{OBryan89}

Furthermore, even if such a $\delta$ fraction of material in the Tetragonal phase existed, it would not be sufficient to produce peaks in an X-Ray diffractogram of comparable magnitude to those coming from the Orthorhombic phase ($>$ 95 \% in mass, since the 7 at.\% of the YBCO would have one Oxygen less per unit cell).
Nor would it explain selective peak appearance.
In addition, if this behavior were intrinsic to the Orthorhombic phase, diffraction patterns for both Orthorhombic cases should be similar in peak positions and relative intensities.\cite{OBryan89}
Although such behavior has been observed previously - three peaks in some cases, or only two flanking peaks in others - it has not been discussed within this framework.\cite{OBryan89}

On the other hand, previous literature has suggested that passage through different superstructures at the Orthorhombic phase may influence the final ordering of Oxygen atoms in the Cu-O chains.\cite{Aligia93,Aligia92PC190}
In addition, measurements of neutron diffraction have shown that different superstructures exhibit not only different occupancies ($\delta$) but also specific characteristics for each one.\cite{Zimmermann03,Ceder90}
Although all superstructures (out of Ortho-I) present a two dimensional interaction, the Ortho-II superstructure has been reported as the only one exhibiting long-range interaction.
Based on the present results, and considering these statements and also taking into account that the T-O transition for YBCO occurs across the Ortho-II superstructure during low-temperature oxygenation (T$_O$ $<$ 400 $^o$C),\cite{LuccasArXiv} we propose that the presence of the Ortho-II superstructure in the path to the final configuration of the Orthorhombic phase is imprinting a lasting signature that is detectable even with commercial equipment.

We propose that this imprint is the result of the Ortho-II superstructure forcing the configuration of Oxygen (vacancies) sites in the Cu-O chains at the final configuration of the material after annealing.
This proposal is in accordance with previous studies that have shown that modulation of Oxygen vacancies within CuO$_2$ planes may cause modulations in charge distribution.\cite{Islam02}
This modulation in charge distribution is strongly related to a modulation in the Cu-O chains (and therefore a modulation in vacancies in their arrangement), which work as a charge reservoir, and potentially influence elastic anisotropy in the material.\cite{BurnsHTc}

\section{Conclusions}\label{SecConcl}

In summary, this work investigates the T-O transition of YBCO at constant temperature for different oxygenation temperatures.
X-Ray diffraction studies reveal differences between samples oxygenated above and below 400 $^o$C.
We propose that passage through the Ortho-II superstructure during low-temperature oxygenation leaves a fingerprint that persists even at high oxygenation levels ($\delta$ $\approx$ 0).

\section{Methods}\label{SecMeth}

Thermogravimetric measurements were performed using a Shimadzu DTG-60H instrument, using alumina ($\alpha$-Al$_2$O$_3$) as the reference material.
YBCO powder of 5 $\mu$m of grain size was prepared as described in Ref. \cite{LuccasArXiv}.
Evolution of m and DTA was studied starting from fully non-oxygenated material ($\delta$ $=$ 1).
X-Ray diffraction measurements were carried out on a PANalytical Empyrean instrument operating in the range 20$^o$ $<$ 2$\theta$ $<$ 120$^o$.

\section{CRediT authorship contribution statement}\label{SecCred}

RFL: Conceptualization, data curation, formal analysis, investigation, methodology, project administration, resources (YBCO powder samples), supervision, validation and visualization for all the stages involved in this research, as well as writing (original draft).
LG: Conceptualization in DTA-TG experiments, data curation in DTA-TG experiments, formal analysis in DTA-TG experiments, investigation in DTA-TG experiments and resources (YBCO powder samples).
CES: Data curation in X-Ray experiments, formal analysis in X-Ray experiments, investigation in X-Ray experiments, resources (YBCO powder samples) and visualization in X-Ray data.
JAM: Conceptualization in DTA-TG experiments and funding acquisition.
All authors contribute to the review and editing of this manuscript.

\backmatter

\bmhead{Acknowledgements}\label{SecTks}

This work was partially supported by MINCyT through the PICT-2017-2898 and PICT-2020-3758 FONCyT projects, and by CONICET through the PIP-2020-2383 project.
RFL thanks to the Organizing Committee of the 30th International Conference on Low Temperature Physics - LT30.
Authors thank to Technical and Administrative staff of the Instituto de F\'{\i}sica Rosario for constant support.
This work was supported by the Argentinian scientific system against the efforts of current Argentinian government.\cite{TheyWantTheEnd}



\bibliography{Luccas_Ortho-II-Fingerprint}

\end{document}